# From Dye Sensitized to Perovskite Solar Cells, The Missing Link

So-Min Yoo,[1,2] Seog Joon Yoon,[2,3] Juan A. Anta,[4] Hyo Joong Lee,[1,5*] Pablo P. Boix,[6*] and Iván Mora-Seró[2*]

[1] Department of Chemistry, Chonbuk National University, Jeonju, 561–756 South Korea

[2] Institute of Advanced Materials (INAM), University Jaume I, Avenida de Vicent Sos Baynat, s/n, 12071 Castelló de la Plana, Spain

[3] Department of Chemistry, College of Natural Science, Yeungnam University, 280 Daehak-Ro, Gyeongsan, Gyeongbuk 38541, Republic of Korea

[4] Área de Química Física, Universidad Pablo de Olavide, Sevilla, Spain

[5] Department of Bioactive Material Sciences, Chonbuk National University, Jeonju, 561–756 South Korea

[6] Institut de Ciència Molecular, Universidad de València, C/J. Beltran 2, Paterna, Spain

* Corresponding author: solarlee@jbnu.ac.kr, pablo.p.boix@uv.es, sero@uji.es

**Abstract**

Fundamental working mechanisms of perovskite solar cells remain an elusive topic of research. Impedance Spectroscopy has been key to characterize the main physical processes governing the behavior of dye and quantum dot sensitized solar cells, which are considered, in many aspects, predecessor technologies. In contrast, IS application to perovskite-based devices generates uncommon features and misleading outputs, mainly due to the lack of a stablished model for the interpretation of the results. In this work we control the perovskite precursor concentration to fabricate a series of perovskite-based solar cells with different amounts of perovskite absorber. Low concentration devices present the well-known dye sensitized solar cell (DSSCs) impedance pattern. As the amount of perovskite is increased, the characteristic impedance spectra of thin-film perovskite solar cells (PSCs) arises. This transition is characterized by a change in the working principles, determined by an evolution of the dominant capacitance: from the intermediate frequency chemical capacitance of TiO$_2$ in devices with isolated perovskite domains, to a large low-frequency capacitance signal which divides the spectra in two sections, yet with no direct influence in final device performance. This study allows to link experimentally, in terms of impedance behavior, PSCs with the rest of solar cell devices via DSSCs. We observe that it is not possible to assign a single physical origin to the different resistances determined in the impedance spectra except for the series resistance. In contrast, resistive element present contributions from different physical processes, observing a transport-recombination coupling. Based on this analysis we provide an equivalent circuit model to evaluate the impedance pattern of PSCs in terms of the processes directly affecting the final performance (i.e. considering transport-related and recombination-related losses), a crucial tool for further development of perovskite photovoltaics.

**Introduction**

More than a decade ago, a new family of light sensitizers, the quantum dots (QDs), experienced an increase of performance and popularity among the dye sensitized solar cell (DSSC) community.[1] DSSCs themselves embodied a new concept[2] on which light abortion and charge transport processes occur in different media, in contrast with previous silicon and semiconductor thin film solar cells. This characteristic relaxes the material quality requirements, enabling high performance solar cells at lower fabrication costs. Intensive research targeted a thorough understanding of the DSSCs working principles,[3-5] the search of better performing materials[5] and the upscaling of the cell production.[6] In this context, the use of semiconductor QDs with nanometer size as sensitizers was very appealing.[1,7] Due to their size, they can be embedded in the mesoporous wide bandgap semiconductor layer acting as electron transporting material (ETM), and similar working mechanisms as in DSSCs were found.[8] QDs can be synthesized with very simple techniques such as chemical bath deposition, successive ionic layer adsorption and reaction, or in colloidal form by hot injection,[7] which permitted multiple groups around the world to initiate research on solar cells with undemanding equipment and facilities, spreading the research interest on this topic.

Since the first reports, the photoconversion efficiency of QD sensitized solar cells (QDSSCs) has continuously increased up to current values close to 13%.[7] The most popular QDs employed for sensitization were chalcogenides semiconductors. However, in 2009 Miyasaka and coworkers[9] reported a new family of quantum dot materials, the halide perovskites, with performances on par with the best chalcogenides-based cells at that time. A couple of years later, Park and coworkers reported 6.5% efficiency with QDSSCs using halide perovskite as a sensitizer, again with record performances.[10] Note that in these cases there was no evidence about quantum-confined size effects and consequently nanoparticles (NPs) appear to be a more appropriate term than perovskite QDs to describe the perovskite nature. However, there is no significant effect in the solar cell working principles between NPs and QDs and consequently we refer to these devices with the most extended name of QDSSCs.

Despite the very competitive efficiency reported for halide perovskite sensitizer in liquid QDSSCs, these materials did not receive much attention due to their instability in liquid iodine redox electrolyte. However, the use of a solid hole transporting material (HTM) in 2012[11, 12] circumvented this problem and significantly increased the efficiency, which transfigured the photovoltaic field.[13] Nowadays, perovskite solar cells (PSCs) have reached efficiencies surpassing 24%,[14] higher than all previously developed thin-film technologies and on pair with those of silicon solar cells. Their continuous performance increase has been the fastest ever reported for any photovoltaic technology,[13] yet with the understanding of the device mechanisms lagging behind in many aspects.

PSCs have shown unconventional behaviors, such as hysteresis of current-voltage (J-V) curve,[15] preconditioning effect,[16, 17] negative capacitance[18-20] or high capacitance at low frequency.[21, 22] Most of these effects have been associated, in different ways, to the soft nature of halide perovskite lattice, which allows for ion motion under light[23] and applied bias.[24] In this context, impedance spectroscopy (IS) should be a valuable asset, as it is a non-destructive technique that allows the solar cell characterization at different working conditions of illumination and applied bias.[25] IS is a frequency domain technique that permits decoupling physical processes occurring at different time scales. This feature makes IS a very appreciated solar cell characterization tool for experimentalists,[26-28] as it can distinguish and characterize different physical processes, including the ones directly affecting the device performance such as charge transport, transfer and recombination.[25] It has been particularly valuable for DSSCs, were IS allows, in a single set of measurements, the determination of ETM resistivity, the counter electrode contribution to the series resistance, the diffusion of the redox species in the electrolyte, and the recombination and the band alignment contribution to the obtained open circuit potential, $V_{oc}$.[25] All these advances were attained by the development of a model which considered the different physical processes occurring in DSSCs, from which an equivalent circuit was derived and subsequently applied to the fitting of the experimental impedance spectra.[3, 29, 30]

The success story of IS' application to DSSC resulted in its almost immediate use to study PSC,[19, 31-34] also with comparative studies between both kinds of cells.[22, 35] However, a univocal interpretation of the PSC IS pattern has been elusive and a universal model capable of describing both types of cells is still missing. In contrast to DSSCs, in which the variation of a single component of the solar cell (such as mesoporous layer thickness, redox couple, counter electrode, etc) affects mainly a specific section of the impedance pattern, these changes affect the whole PSC IS spectrum, hindering the association of the circuit elements to concrete physical processes. In this work, we track the evolution of impedance spectra from the well-known DSSC to a PSC configuration by varying the concentration of chemical precursors for the perovskite formation. The capacitive signal variations allow us to identify the main features of the impedance pattern. These are employed to propose a rationalized model for perovskite solar cells' impedance response, focusing on the transport and recombination processes that have a direct relation with the final device performance. As a result, we provide a helpful tool for the performance characterization to contribute to further improvement of this technology.

**Results**

The analyzed samples were prepared on transparent conducting glass substrates, $SnO_2$:F (FTO), depositing a compact $TiO_2$ layer and a $TiO_2$ mesoporous layer (~400 nm thick), as it is conventionally used in DSSC as well as in some PSC. $CH_3NH_3PbI_3$ ($MAPbI_3$), and $CsPbI_{1.3}Br_{1.7}$ were deposited on these films by sequential deposition of precursors, known as 2 steps deposition method (details in experimental section). Different hybrid and inorganic perovskite materials were used in order to highlight the generality of the approach as the same trends have been observed in all cases. Figure 1a displays a picture of the electrodes sensitized with $MAPbI_3$ using different $PbI_2$ concentrations and a fixed 0.1 M concentration of MAI, see experimental section for further details. The final amount of deposited perovskite increases with the $PbI_2$ concentration, as confirms the visual darkening of the sensitized electrodes, see Figure 1a. Cross-sectional SEM images of some of the prepared samples, Figure 1b-g, show how an increasing amount of $MAPbI_3$ is preferentially deposited inside $TiO_2$ mesoporous scaffold as the PbI$_2$ concentration increases. Top views of sensitized electrodes are depicted in Figure S1. Samples with 0.1 M PbI$_2$, Figure 1b, are indeed impossible to differentiate from bare TiO$_2$ ones, and the deposition results in unconnected <10 nm MAPbI$_3$ nanoparticles, as previously reported.[36] Samples deposited with 0.2 of PbI$_2$, Figure 1a, look very similar, with the TEM analysis confirming the formation of small (< 10 nm) isolated clusters of perovskite on the TiO$_2$ surface, see Figure S2. However, a close inspection of samples with 0.3 M PbI$_2$, Figure 1c, reveals small regions covered with perovskite where the nanoparticles have grown enough to merge and coat small parts of the mesoporous layer. From 0.5 M PbI$_2$, Figure 1d, larger perovskite covered sections are evident, with higher preferential deposition of perovskite close to the compact TiO$_2$ / mesoporous TiO$_2$ interface. If PbI$_2$ precursor concentration is further increased to 0.7 M, Figure 1e, the deposited perovskite practically fills all the voids of the mesoporous layer, with a just slight covering of the upper part of the scaffold layer. For 0.9 M PbI$_2$, Figure 1f, perovskite islands on top of the mesoporous layer can be identified, and these structures are further incremented for 1.1 M PbI$_2$, Figure 1g and S1.

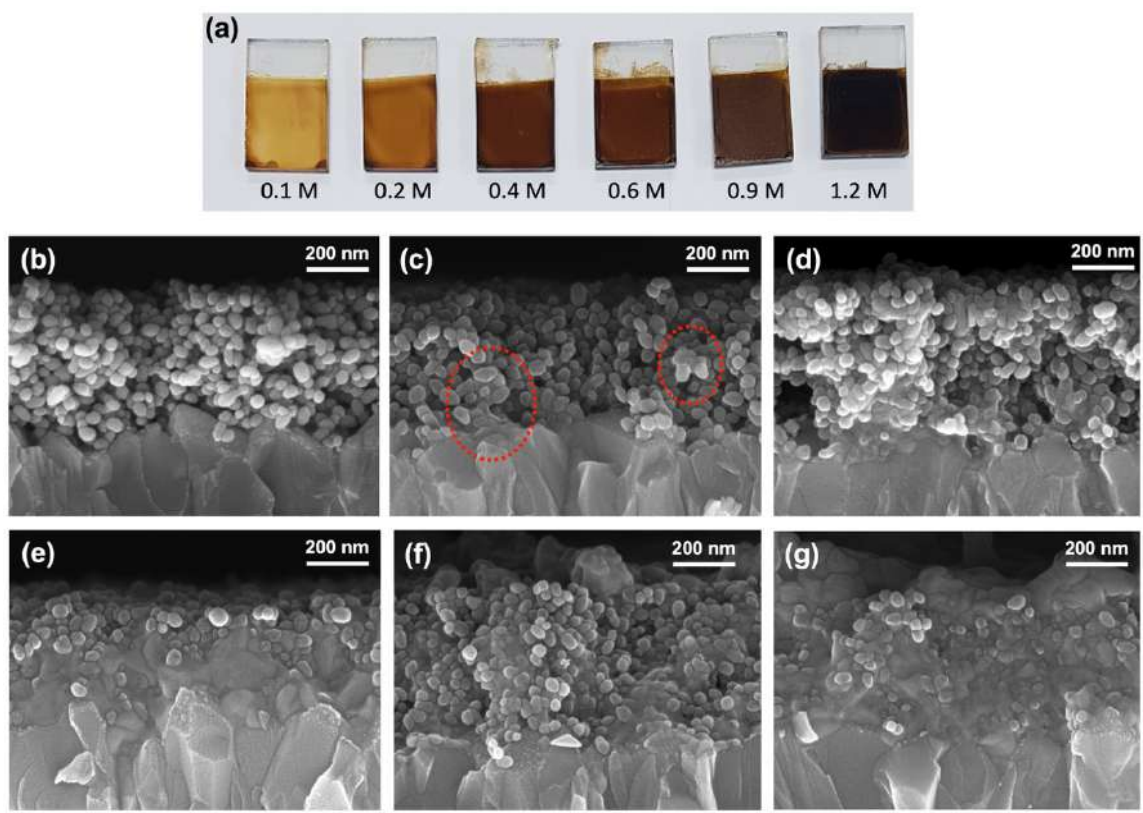

**Figure 1. (a)** Picture of the halide perovskite MAPbI$_3$ sensitized electrodes. SEM micrographs of the cross section of sensitized electrodes using as precursor concentration 0.1 M MAI and PbI$_2$ of **(b)** 0.1 M; **(c)** 0.3 M, dotted lines encircle regions where MAPbI$_3$ deposits are visible at this magnification; **(d)** 0.5 M; **(e)** 0.7 M; **(f)** 0.9 M and **(g)** 1.1 M.

These TiO$_2$/MAPbI$_3$ electrodes can be implemented in functional solar cells by the subsequent deposition of spiro-OMeTAD as hole transporting material (HTM) followed by evaporated Au as charge extracting contact. The forward J-V curves obtained for devices prepared with bare TiO$_2$ electrode with no sensitizer, and MAPbI$_3$ deposited with different PbI$_2$ concentrations are plotted in Figure 2a. The respective solar cell parameters are collected in Table S1. As expected and previously reported,[36] the performances are lower than the state-of-the-art for PSCs, mainly due to the small short circuit currents, $J_{sc}$. Note that optimal standard configuration of PSCs requires a thinner mesoporous layer and a thicker perovskite overlayer on top of the scaffold.[37] Bare TiO$_2$ device results in a very low $J_{sc}$, as just high energy UV photons can be absorbed. As the perovskite amount in the TiO$_2$ increases with the PbI$_2$ concentration, the semitransparent solar cells performance is enhanced, the trend changes after 0.7M. The hysteresis behavior of representative J-V curves is plotted in Figure S3. The analyzed samples do not present significant hysteresis, independently of the precursor concentration used in their fabrication.

Impedance spectroscopy measurements were carried out for different samples at 0.1 sun illumination at varying applied forward bias. Limited light intensity ensures device stability during IS measurements, while it still reproduces the characteristic behavior of PSC.[38] We have verified the stability of the devices by measuring at maximum power point during one hour with no reduction of the device performance, see Figure S3d. This is the double of the required time for IS measurement, and also at steady-state conditions. One of the key aspects to identify the physical processes behind the impedance spectrum is the capacitance. The Bode plot of the real part of the capacitance for different concentration of MAPbI$_3$-sensitized cells and bare TiO$_2$ samples is displayed in Figure 2b. For a bare TiO$_2$ sample, two plateaus can be observed in the Bode plot, one at high frequency (HF), ~10$^4$-10$^6$ Hz, and the other at intermediate frequency (IF), ~10-10$^4$ Hz, and low frequency (LF), ~10$^{-1}$-10 Hz. The HF capacitance, $C_{HF}$, is associated with the geometric capacitance, $C_g$,[39] while the LF capacitance, $C_{LF}$, is ascribed to chemical capacitance, $C_\mu$.[30] The former, $C_g$, is the classical electrostatic capacitance produced by the electrical field between two plates, the two contacts in our case. In contrast, $C_\mu$ reflects the capability of a system to accept or release additional carriers due to a change in their chemical potential.[30] Accordingly, $C_\mu$ in a DSSCs reflects the occupation of the electronic states in the mesoporous TiO$_2$ layer due to the injection of the electrons photogenerated in the sensitizer.[3, 30, 40] As a result, it is determined by the ETM, rather than by the sensitizer. The perovskite sensitized device prepared with the lowest PbI$_2$ concentration, 0.1 M, presents exactly the same pattern as the bare TiO$_2$, which indicates that these samples present a classical DSSC behavior. Perovskite nanoparticles, see Figure S2, sensitize the TiO$_2$ ETM, injecting the photogenerated electrons as unveils the observation of the same $C_\mu$ than in bare TiO$_2$ samples.

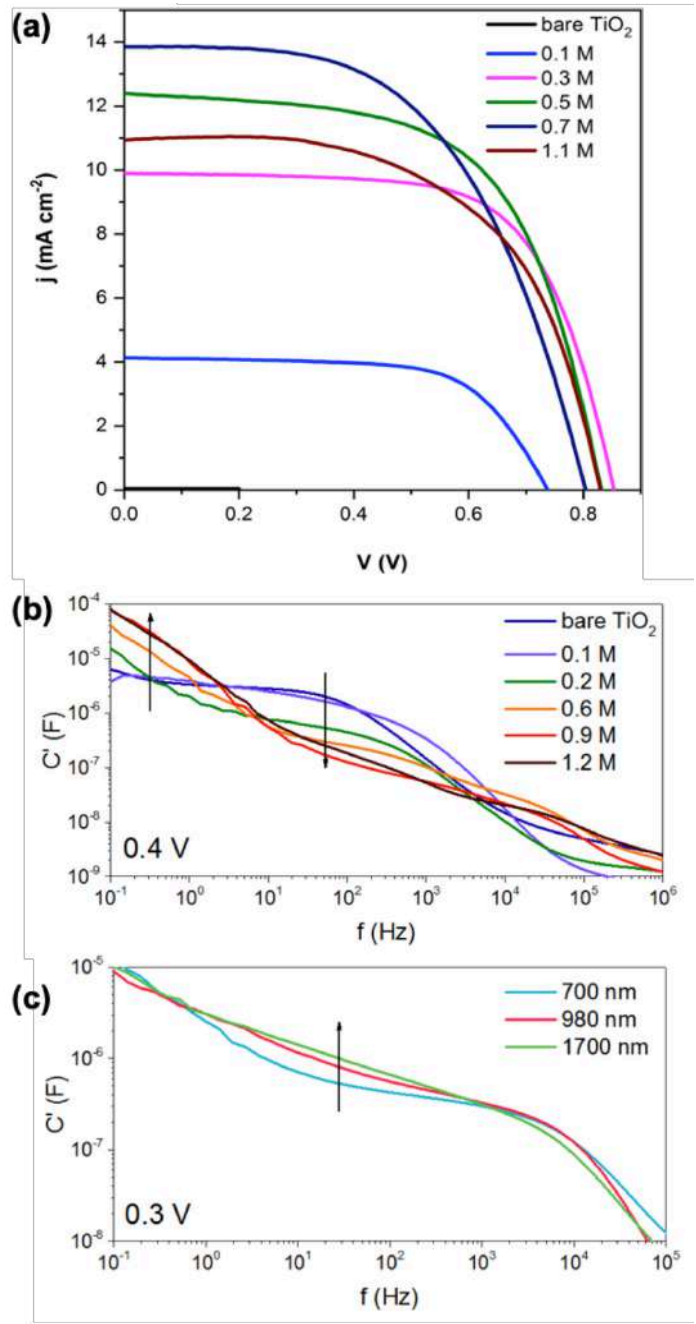

**Figure 2. (a)** J-V curve of MAPbI$_3$ solar cells prepared with precursor concentration of MAI 0.1 M and different PbI$_2$ concentrations. **(b)** Bode plot of the real part of the capacitance of MAPbI$_3$ solar cells prepared on 400 nm thick TiO$_2$ mesoporous layer substrates with precursor concentration of MAI 0.1 M

and different PbI$_2$ concentrations. Arrows indicate the evolution as the PbI$_2$ concentration increases. **(c)** Bode plot of the real part of the capacitance of CsPbI$_{1.3}$Br$_{1.7}$ solar cells prepared on different thickness TiO$_2$ mesoporous layer substrates with precursor concentrations of 0.06 M of CsBr and 0.3 M of PbI$_2$. Arrow indicates the evolution as the thickness of the mesoporous layer increases.

When the amount of deposited perovskite is increased, a progressive decrease in the $C_\mu$ at intermediate frequencies is observed, see Figure 2b. The increase of deposited perovskite creates more percolation pathways, which enable the transport of electrons photogenerated in perovskite to the compact TiO$_2$ layer without being injected into the mesoporous TiO$_2$. Therefore, $C_\mu$ decreases as a result of the lower electronic states occupation in the mesoporous TiO$_2$ scaffold.

Concurrently, a new capacitive feature appears at LF, which progressively increases up to a saturation value for the samples with the larger amounts of deposited perovskite, see Figure 2b. Note that the behavior of the capacitances observed at IF and LF is clearly different for samples with different TiO$_2$ mesoporous layer thicknesses, see Figure 2c. The IF capacitance increases with the mesoporous layer thickness, a characteristic signature of $C_\mu$, as the density of states increases with the material volume.[30] In contrast, the observed capacitance at LF does not exhibit this behavior, and shows similar values for all three-layer thicknesses, pointing to an interfacial process. The different weight of these capacitive processes will influence the IS pattern, not just in the Bode plots in Figures 2b and 2c, but also the Nyquist plots in Figure 3.

In Figure 3 we track the evolution of the IS of the fabricated devices from those presenting typical DSSC behavior, prepared with a low amount of perovskite, to those with a thin-film PSCs configuration, prepared with higher precursor concentration, see Figure 1. Figure 3a displays the Nyquist plot of a MAPbI$_3$ solar cell using 0.2 M PbI$_2$ concentration. As we discussed in the analysis of Figure 2b, the devices fabricated with low precursor concentration behave as standard all-solid DSSC. As in DSSC, the IS pattern depends on the applied bias. We have selected 0.6 V forward bias in Figure 3a as it displays the richest impedance pattern, with two arcs, at HF and LF, connected by a straight line at IF with slope close to 1. Nyquist plots of the same at other applied bias are represented in Figure S4. This pattern can be easily fitted with the standard equivalent circuit previously developed for all-solid DSSCs,[40] based on a transmission line (TL) circuit developed by Bisquert and coworkers[29, 40] and plotted in Figure 3g, in parallel to the geometric capacitance, $C_g$. The TL considers the transport of both carriers across the active layer, coupled with the recombination processes and the occupation of the mesoporous $TiO_2$ electronic states. These physical processes are represented by the different elements of the equivalent circuit, Figure 3g, and are denoted in lower case letters as they are distributed elements along the TL.[3, 29] $C_\mu = c_\mu \cdot L$, where L is the active layer thickness, representing the chemical capacitance. $R_{rec} = r_{rec}/L$ is the recombination resistance inversely proportional to the recombination rate.[3, 25, 29] $R_{te} = r_{te} \cdot L$ is the electron transport resistance, while $R_{th} = r_{th} \cdot L$ is the hole transport resistance. In liquid DSSCs, hole transport takes place by ionic diffusion through the electrolyte, with a low resistance associated and consequently $R_{th}$ is neglected.[3] In contrast, when the liquid electrolyte is replaced by a solid HTM, higher values of $R_{th}$ were observed as it is featured in the spectra.[40] This equivalent circuit was used to simulate the IS pattern of all-solid DSSCs, see Figure 3d,[40] and also allows obtaining the IS pattern observed for the fabricated devices with low precursor amount of perovskite presenting DSSC behavior, see Figure 3a. TL equivalent, Figure 3g, can produce two different patterns depending on the diffusion length, for diffusion lengths longer than the layer thickness (transport resistances smaller than recombination resistance), the diffusion-recombination impedance is obtained, while for diffusion lengths shorter than the layer thickness (transport resistances larger than recombination resistance), the Gerischer impedance is observed, see Figure S5.[41] The pattern in Figure 3a, where the prolongation of the straight line cuts the LF arc, corresponds to a diffusion-recombination impedance. Hereafter we will consider just this case if nothing else is mentioned.

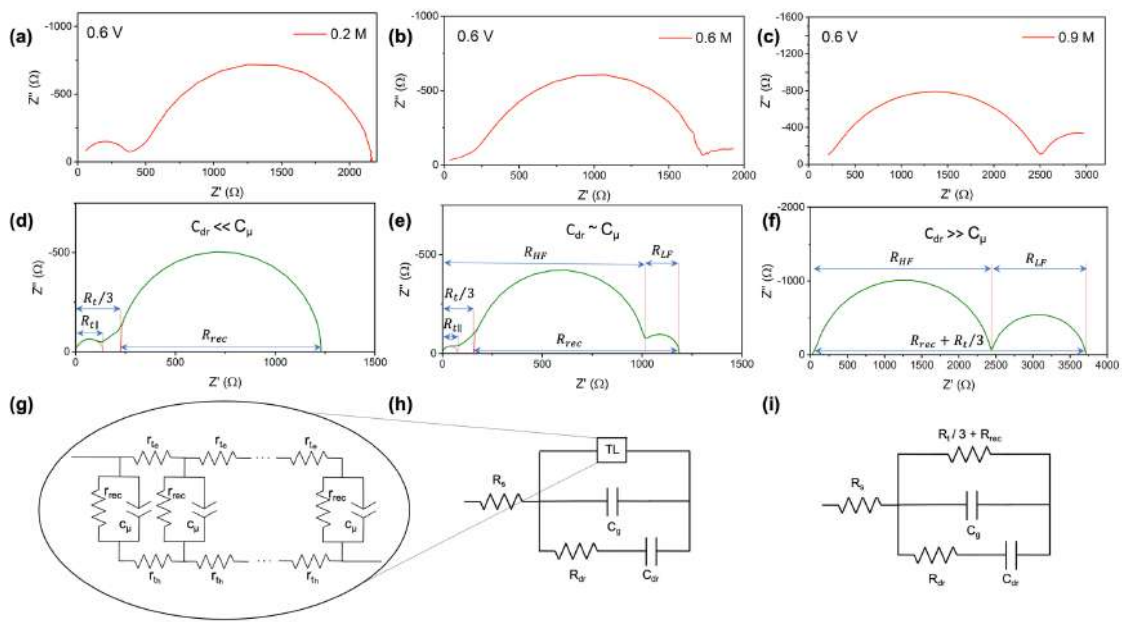

**Figure 3.** Nyquist plots at 0.6 V forward applied bias under 0.1 sun illumination of MAPbI₃ solar cell prepared with 0.1 M of MAI and a PbI₂ concentration of **(a)** 0.2 M; **(b)** 0.6 M **(c)** 0.9 M. **(d)** Simulated Nyquist plots using the **(g)** transmission line equivalent circuit with transport resistance for both electrons and holes as developed for all-solid DSSCs.[40] Parameters, (from general model in h): $R_s$=25 Ω, $R_{te}$=70 Ω, $R_{th}$=500 Ω, $R_{rec}$=1000 Ω, $C\mu$=10⁻⁵ F, $C_g$=10⁻⁸ F, dielectric relaxation(dr)-like resistance $R_{dr}$=7000 Ω, and capacitance $C_{dr}$=10⁻⁶ F. **(e)** Simulated Nyquist plots using the equivalent circuit **(h)** including the transmission line (in (g)) with a dielectric relaxation like branch characterized by a dielectric relaxation resistance, $R_{dr}$, and a dielectric relaxation capacitance, $C_{dr}$.[40] Parameters: $R_s$=25 Ω, $R_{te}$=70 Ω, $R_{th}$=500 Ω, $R_{rec}$=1000 Ω, $C\mu$=10⁻⁶ F, $C_g$=10⁻⁸ F, $R_{dr}$=7000 Ω, and $C_{dr}$=10⁻⁵ F. **(f)** Simulated Nyquist plots using **(i)** a simplified equivalent circuit with respect the one depicted in in (h) considering the case $C_g$, $C_{dr} \gg C_\mu$ Parameters (from general model in h): $R_s$=25 Ω, $R_{te}$=70 Ω, $R_{th}$=500 Ω, $R_{rec}$=3500 Ω, $C\mu$=10⁻⁷ F, $C_g$=10⁻⁷ F, $R_{dr}$=7000 Ω, and $C_{dr}$=10⁻⁴ F.

The features observed in the impedance spectra of an all-solid DSSC, Figure 3a, can be correlated to the different elements of the equivalent circuit of a TL with an additional $C_g$ in parallel. The HF arc, see Figure 3d, is originated by the coupling of $C_g$, that presents a lower value than $C_\mu$, with the parallel association of $R_{te}$ and $R_{th}$:[42]

$$R_{t\parallel} = \frac{R_{te} \cdot R_{th}}{R_{te} + R_{th}} \quad (1)$$

Consequently, the characteristic time of the HF arc considering only the elements of the TL is $\tau_{HF-TL} = R_{t\parallel} \cdot C_g$. This arc is followed by a straight line. The length of the HF arc plus the straight line in the Nyquist plot is one third of the total transport resistance, $R_t$, obtained as the series association of electron and hole transport resistances:

$$R_t = R_{te} + R_{th} \quad (2)$$

Ultimately, the LF arc is determined by the coupling of $C_\mu$ with $R_{rec}$, and the resistance at the low frequency limit (DC limit) is $R_{rec}+R_t/3$. Note that for the sake of clarity in Figure 3g, a simplified TL equivalent circuit is represented, as previously used to describe the all-solid DSSCs.[40] A more detailed discussion of the ambipolar impedance equivalent circuit can be found elsewhere.[43]

Consequently, the fabricated samples with low amount of perovskite present exactly the same behavior than standard all-solid DSSCs. However, when the amount of deposited perovskite is increased, the resulting impedance spectra evolves towards a different pattern. In Figure 3b the IS spectra of a device fabricated with higher concentration of PbI₂, 0.6 M is represented. These conditions also result in a pattern similar to that of DSSCs, Figure 3a, yet with an additional arc at LF. The presence of this new arc is due to the appearance of the new capacitance effect at low frequencies, observed in Figure 2b. In this sense, this LF feature is the missing-link that connects DSSCs to PSCs, and it is related to the high LF capacitive effect observed at LF in PSCs under illumination.[21] The straight line observed in the pattern presents the transport resistance fingerprint, Eq. (2), but the smaller HF arc indicates that one of the carrier transport resistances has diminished, Eq. (1). This can be attributed to a $R_{te}$ reduction due to the new pathways created for electrons though the perovskite, also reflected in the concurrent chemical capacitance decrease, Figure 2b. The full impedance spectra cannot be fitted with the parallel association of TL and $C_g$, as in Figure 3a. We propose the incorporation of a new parallel branch, as depicted in Figure 3h. This branch presents a behavior similar to a dielectric relaxation that can follow frequency changes at LF but not at HF. We tentatively name this new branch a dielectric relaxation (dr)-like one, due to this analogous behavior, yet without attributing the exact physical origin. It is composed by a dr-resistance, $R_{dr}$, and a dr-capacitance, $C_{dr}$, which are associated to the perovskite active layer and that encompasses the low frequency capacitive effect observed in Figure 2b. A series resistance, $R_s$, is also added to the extended TL model for PSCs to incorporate the electrodes and wiring resistance, Figure 3h.

**Discussion**

The evolution from DSSC behavior to a more complex one when the perovskite amount increases in the device, (description of this effect in terms of the equivalent circuits depicted in Figure 3g and 3h) provide an adequate framework for a deeper discussion. Multiple analyses can be performed around the different concepts introduced in the previous section. However, the main interest of our work is to connect the impedance behavior of PSCs with the general impedance behavior of solar cells, and leveraging this knowledge to provide an easy to tool to evaluate the PSCs performance in terms of the impedance spectra.

With the analysis performed for 3a and 3b patterns, though the incorporation of a dr-brach, we have connected devices with typical PSC responses with devices with DSSCs behavior. Nonetheless, it is very important to highlight that this dr-like branch acts as an open circuit at DC conditions, thus it does not affect directly the J-V curve or the cell performance, but it can still result in hysteretic behavior as analyzed elsewhere[44, 45] and in Figure S3. To this extent, the determination of the exact nature of the physical processes involved in $R_{dr}$ and $C_{dr}$ elements is beyond the scope of this manuscript, which intends to focus on the reproducibility of the experimentally observed impedance patterns and the identification of the transport and recombination processes in PSCs. These processes have a direct effect in the JV curve and consequently in the device performance. In addition, the physical origin of the characteristic behavior of PSCs at LF[21] is already discussed in the literature[44, 46-49] and it is briefly summarized below.

A theoretical approach to the model schematized in Figure 3h was suggested for the general case of solar cells with dielectric relaxation, drift-diffusion transport and recombination, where the electrical field was approximatively introduced in the diffusion-recombination impedance by the addition of the dr-parallel branch.[42] Since the first reports of large, low frequency capacitance under illumination,[21] here represented by $C_{dr}$, this specific PSCs characteristic has attracted great interest. The dielectric relaxation-like behavior designates a physical process that cannot follow the fast electrical field changes at HF. This fact is compatible with an ion migration-based scenario, which can respond to the LF but not to the high ones. Actually, several reports agree in a central role of ions in order to explain this particular observed behavior.[44, 46, 49] These reported models provide different interpretations for the low frequency capacitance, including the accumulation of majority carriers at the $TiO_2$ interface compensated by ions,[22, 46] the interplay between ionic and electronic charge transfer at perovskite-contact interfaces controlling electronic injection and recombination,[47, 49] or the ion-modulated recombination.[33, 44, 48] In the latter works, the LF capacitance is described as an "apparent capacitance" due to the out of phase modulation of either the recombination or the injection current caused by mobile ions in the perovskite film, in contrast with a capacitance directly related with charge accumulation.[46] Note that all these processes which intrinsically cannot follow high frequency variations as well as electrical field effects in drift-diffusion transport and recombination,[42] could be reduced to circuit elements analogous to $R_{dr}$ and $C_{dr}$, which provides a high degree of generality to the model presented here. Most of these works do not analyze the IS in terms of equivalent circuit models, which are effective tools to a broader community. Interestingly, in the work of Moia et al.[49] part of the discussion is in terms of equivalent circuits and similar circuits that the one proposed here could be obtained from a different approach. In our approach, however, we develop the equivalent circuit from the experimental data and incorporate the effect of charge transport processes, which can strongly affect the device performance as we discuss below.

Coming back to the analysis of the IS, note that the standard TL for all-solid DSSCs, Figure 3g, is a particular case of the extended model, Figure 3h, in which $C_{dr} \ll C_\mu$ (see the simulation parameters in figure caption of Figure 3d and 3e). The effect of the dr-like branch is the splitting of the diffusion-recombination TL pattern into two parts, $R_{HF}$ and $R_{LF}$ at HF and LF, see Figure 3e for a forward applied bias of 0.6 V and Figure S6 for the impedance spectra at other applied bias. Due to the diffusion-recombination pattern of TL, $R_{HF}$ and $R_{LF}$ can be expressed as:[42]

$$R_{HF} = \frac{(R_{rec}+R_t/3) \cdot R_{dr}}{(R_{rec}+R_t/3)+R_{dr}} \quad (3)$$

$$R_{LF} = (R_{rec} + R_t/3) - R_{HF} \quad (4)$$

The size of each of the two parts depends on the ratio:

$$\frac{R_{HF}}{R_{LF}} = \frac{R_{dr}}{R_{rec}+R_t/3} \quad (5)$$

The DC resistance, $R_{DC}$, which can be used to reconstruct the JV curve, can be extracted from the low frequency limit:

$$R_{DC} = R_s + R_{rec} + R_t/3 \quad (6)$$

Note that $R_{DC}$, which determines the JV curve, does not include any of the parameters of the relaxation-like branch as previously commented.

The pattern observed in Figure 3b can be considered a transition pattern from DSSCs to PSCs. Finally, if the amount of perovskite is further increased (> 0.6 M), the TL pattern is lost and the impedance spectra exhibits just two arcs at high and low frequencies, see Figure 3c. The impedance spectra of the same sample at different applied bias is depicted in Figure S7. The transition from a TL patter to just two arcs is driven by the $C_\mu$ reduction and $C_{dr}$ increase, as previously discussed and displayed in Figure 2b. When the $C_\mu$ becomes lower than $C_{dr}$ and $C_g$, the extended TL equivalent circuit, Figure 3h, can be simplified to the circuit depicted in Figure 3i. A similar equivalent circuit was used by Pascoe et al. to analyze planar devices, but it was not derived from a most sophisticated general model[32] and further implications of transport, that we discuss below, were not addressed.[32, 42] This circuit generates two arcs in the impedance spectra, see Figure 3f, where the high and low frequency resistances, $R_{HF}$ and $R_{LF}$, are defined by Eqs. (3) and (4) respectively. The capacitance Bode plot corresponding to the data in Figure 3c and S7 is displayed in Figure S8 and represents the characteristic behavior of this kind of devices. At HF $C_g$ is the dominating feature, while capacitance increases to $C_{dr}$ at LF. As previously discussed, the significant decline of $C_\mu$ results from the reduction of the electron injection into TiO$_2$ scaffold when the system get close to a PSC configuration, as electrons are transported through the low resistivity perovskite rather than through the TiO$_2$.[50] Besides, halide perovskites are characterized by a low density of states below the bandgap,[51] which explains the much lower $C_\mu$ than for TiO$_2$, and makes $C_\mu$ negligible in comparison with $C_g$ and $C_{dr}$. Pocket et al.[33] provided a theoretical estimation of the chemical capacitance in a planar perovskite, which is of around $10^{-8}$ F/cm$^2$ for an open circuit voltage of 1V. This value is lower than the observed for $C_g$ and $C_{dr}$.

The impedance spectra observed in Figure 3c is the widely observed pattern in PSCs,[32, 33, 39, 46, 48, 52-54] including the high performance devices.[39, 55, 56] Let us discuss the implications of the model for the understanding of the PSCs, especially regarding transport and recombination processes. The double arc spectra observed in Figure 3c allows only the determination of two resistances, $R_{HF}$ and $R_{LF}$, as classically carried out in the literature.[33, 46] Consequently, it is not possible to obtain simultaneously $R_{rec}$, $R_t$ and $R_{dr}$. In addition, $R_{rec}$ and $R_t$ are coupled in the determination of $R_{DC}$, see Eq. (6). This is the origin of the confusing results obtained from the IS of PSCs. In the case of high-performance PSC, it can be considered that $R_t \ll R_{rec}$, thus $R_t$ can be neglected in the Figure 3i equivalent circuit. In this case, $R_{rec}$ can be simply obtained by the addition of $R_{HF}$ and $R_{LF}$.

Most of the previous studies in impedance spectroscopy of PSCs have been made in terms of $R_{HF}$ and $R_{LF}$, but an important conclusion of this work is that $R_{HF}$ and $R_{LF}$ are not related to a single physical process but to a combination of charge transport, recombination and dielectric relaxation-like processes, as defined by Eq. (3) and (4). The connection the results reported in this work with some of the previous IS characterization results[25, 33, 44, 46-49, 52, 53, 57, 58] is discussed in supporting information, Note 1.

Finally, an important factor in solar cell performance is the effect of the contacts and interfaces that, beyond $R_s$, has not been explicitly considered in the discussed model so far. It has been observed that an increase of the transport layers' thicknesses, and consequently contact transport resistances, does not change the IS pattern shape.[39, 59] Basically, two arcs without any additional feature are still observed, but their size varies with the thickness of both ETM[39] and HTM[59], as seen in Figure S9. In the example discussed in Figure S9, the association of $R_{rec}$ with $R_{HF}$, $R_{LF}$, or $R_{HF}+R_{LF}$, can result in a misleading conclusion, as an increase of the HTM thickness produces an apparent enhancement of $R_{rec}$ (i.e. reduction of recombination rate), but at the same time a decrease of $V_{oc}$ is also observed. This contradiction is solved by the model devised here, as the increase of $R_t$ at the contact layers causes the $R_{HF}$ and $R_{LF}$ increase, see Eqs. (3) and (4), unrelated to a recombination reduction. Consequently, $R_t$ in Figure 3i can also contain the transport resistance at the contact layers, which has very important implications for the interpretation of the IS pattern.

The large variety of PSC materials and configurations have also generated a significant amount of particular cases that provide richer perovskite patterns than the double arc spectra of Figure 3c.[18, 39] A systematic characterization will be needed to incorporate these features in the general model here proposed. In this context, we would like to use this discussion to provide some indications that should be considered in this future work. For instance, the device interfaces can also influence the IS pattern, allowing the observation of and additional feature at intermediate frequencies.[18, 39, 60] An interface can be characterized by a resistive element, $R_{interfaces}$, if an energy lost is produced by the charge transfer at the interface, in parallel with an interfacial capacitance.[61] If interfacial capacitance, $C_{interfaces}$ present higher value than $C_g$ an IF feature, as observed,[18, 39, 60] will be obtained. On the contrary if $C_{interfaces} < C_g$ just a couple of arcs would be observed $R_{interfaces}$ should produce an analogous effect than transport resistance in the equivalent circuit of Figure 3i. In this case, a parallel RC circuit has to be added in series in the upper branch of Figure 3i. On the other hand, negative capacitance at LF and/or loops at IF have been also observed in the IS of PSCs. These features produce a decrease of $R_{DC}$, and consequently, of the device performance, especially in the case of negative capacitance at LF,[20] and have been observed in different kind of samples.[19, 20, 22, 39, 47, 62, 63] Some, studies associate this negative capacitance to a particular case of the apparent capacitance observed at LF.[47, 49] Their effect in the IS pattern could be simulated by the introduction of an inductor element in parallel with $C_{dr}$.[18, 39, 64]

**Conclusions**

We have tailored the solar cell behavior from DSSC-type to thin-film PSC by the control of the amount of perovskite deposited on mesoporous $TiO_2$ layer. This evolution has been investigated by means of impedance spectroscopy. When the amount of deposited perovskite increases, $C_\mu$ decreases due to lower injection into the mesoporous $TiO_2$ and consequently a lower occupation of electronic states in the mesoporous $TiO_2$. This is caused by the generation of percolation pathways which enable the electron extraction through the perovskite itself. Concurrently, the low frequency capacitance under illumination, characteristic feature of PSCs, increases with the amount of deposited perovskite. Samples with very low perovskite concentration follow the well-known impedance model for all-solid DSSCs with a TL. But as the amount of deposited perovskite increases, i.e. samples configuration is closer to the standard PSC one, this pattern is divided into two parts by the action of a dielectric relaxation-like branch. When the amount of perovskite is further increased and generates an interconnected film of perovskite, the decrease of $C_\mu$ and the increase of $C_{dr}$ results in a simple pattern with two arcs, commonly observed in thin film-type PSCs. In this pattern, transport (including that of the perovskite layer, ETM and HTM) and recombination are coupled, which explains the high degree of confusion of the different IS analysis reported. The dielectric relaxation-like branch does not contribute to the $R_{DC}$ and consequently it does not directly affect the cell performance. This provides a new point of view in several aspects regarding the study of PSCs. On one hand, the IS pattern is experimentally related with the DSSCs connecting PSC with the general group of photovoltaic devices. On the other hand, we have shown that recombination and transport are coupled in the determination of $R_{DC}$ and consequently in the interpretation of JV curve and device performance. Finally, we provide an easy tool, in the form of an equivalent circuit, to evaluate the PSCs performance in terms of the impedance spectra, that will undoubtedly help the experimentalist in the device characterization and optimization processes. This paves the way for an interpretation of the physical processes that can be identified by IS affecting the device performance, a broadly extended characterization technique that can provide important clues about the working principles of perovskite solar cells.

**EXPERIMENTAL SECTION**

*Device Fabrication.* A fluorine-doped tin oxide-coated glass (FTO) substrates (25 x 25 mm, Pilkington TEC15, ~15 Ω/cm) were partially etched with zinc powder and diluted HCl solution. The substrates were cleaned with distilled water with soap and rinsed with distilled water and acetone. Then, the substrates were sonicated sequentially in acetone and ethanol for 15 min and dried with compressed air. Finally, the substrates were treated with UV-O₃ lamp for 15 min. The TiO$_2$ compact layer was deposited by spin-coating with 80 μL of Ti-alkoxide solution (ShareChem, SC-BT060) at 4000 rpm for 30 sec. The films were dried at 150 °C for 10 min and heated gradually up to 500 °C and then kept for another 30 min. Mesoporous $TiO_2$ layer was deposited by spin-coating an aliquot of diluted $TiO_2$ paste (a 1.0 g of paste, SC-HT040 from ShareChem was mixed well with a 5.0 mL of pure ethanol) at 1000 rpm for 20 sec and heated gradually up to 500 °C, which gave about 400 nm thick mesoporous $TiO_2$ film. For different thickness, three different suspensions containing 1.0 g of $TiO_2$ paste (ShareChem, SC-HT040) diluted with 2.0, 3.0, 4.0 mL of pure ethanol each were used to make $TiO_2$ films with thickness of about 1700, 980, and 700 nm, respectively.

To deposit the $CH_3NH_3PbI_3$ (MAPbI$_3$) perovskites onto the meso-$TiO_2$ film by 2-step method, 0.1~1.2 M solutions of lead iodide ($PbI_2$, TCI, 99.999%) in DMF/DMSO (4:1, v/v) and 0.1 M solution of MAI (Dyesol) in 2-propanol were used as each precursor source. A 100 μL of $PbI_2$ precursor solution was dropped over meso-$TiO_2$ film/FTO and spin-coated at 2500 rpm for 1 min and dried at 70 °C for 10 min. Then the films were dipped in a MAI solution for 30 sec, rinsed with 2-propanol, dried by $N_2$ gas, and annealed at 70 °C for 5 min.

As for deposition of $CsPbI_{1.3}Br_{1.7}$ perovskites onto the meso-$TiO_2$ film/FTO by 2-step deposition, 0.30 M solutions of lead(II) iodide ($PbI_2$, TCI, 99.999%) in DMF and 12 mg mL$^{-1}$ cesium bromide (CsBr, Sigma-aldrich, 99.999%) in methanol were applied successively as each precursor source. A 100 μL of $PbI_2$ precursor solution was dropped over meso-$TiO_2$ film/FTO and spin-coated at 1000 rpm for 10 sec and then at 4000 rpm for 30 sec. The as-deposited $PbI_2$ was dried at 70 °C for 10 min. For the addition of CsBr, the $PbI_2$-coated electrode was spinned at 2000 rpm and then 500 μL of CsBr solution was dropped to make $CsPbI_xBr_{3-x}$ perovskites. The deposited film was annealed at 250 °C for 5 min followed by 280 °C for 9 min in a $N_2$ atmosphere.

A 60 μL of spiro-OMeTAD solution was added over the MAPbI$_3$ perovskite films and spin-coated at 2000 rpm for 30 sec after staying for 30 sec on the perovskite surface. The spiro-OMeTAD solution was prepared by dissolving (1) 80.0 mM of spiro-OMeTAD in chlorobenzene, then mixed with (2) 64.0 mM of 4-tert-butylpyridine, (3) 16.0 mM Li$^+$ from a stock solution of 290 mg/mL bis(trifluoromethane)sulfonimide lithium salt (Li-TFSI) in acetonitrile. Finally, 70 nm-thick gold (Au) was thermally evaporated on the top of the device as counter electrode. In the case of $CsPbI_{1.3}Br_{1.7}$, 110 mM of spiro-OMeTAD, 88.0 mM of 4-tert-butylpyridine and 22.0 mM Li$^+$ were used in the same manner as above.

*Characterization.* The PV device performances were analyzed under a standard illuminating condition (AM 1.5, 100 mW cm$^{-2}$) using a solar simulator (ABET Technologies Sun 2000) and the current-voltage curves were measured with a digital source meter (Keithley 2612). The light intensity was calibrated with a NREL-certified Si reference cell. The solar cells were masked with a metal aperture of 0.12 cm$^2$ to define the active area. The TiO$_2$/perovskite film morphologies were investigated using a scanning electron microscope (SUPRA 40VP, Carl Zeiss) and EDXS maps were recorded using an Apollo X (EDAX) to define the compositional ratio of I/Br in CsPbI$_x$Br$_{3-x}$. The powdery TiO$_2$/MAPbI$_3$ was collected by scraping the film into as small a piece as possible, and high-resolution transmission electron microscopy (TEM) imaging was carried out using a JEOL (JEM-2010) microscope operated at 200kV.

The impedance spectroscopy measurements were carried out on a PGSTAT30 potentiostat (Autolab). A Xe lamp was used to illuminate the perovskite solar cells and a neutral-density filter was used to adjust the light intensity to 0.10 Sun power. All the impedance spectra were measured ranging between 1 MHz and 0.1 Hz.

**Acknowledgment**

This work has been partially supported by Generalitat Valenciana (SEJI2017/2017/012), Ministerio de Economía y Competitividad of Spain (MAT2017-88905-P, MAT2016-76892-C3-2-R and Red de Excelencia "Emerging photovoltaic Technologies"), the National Research Foundation of Korea (NRF-2017R1D1A1B03028570) and the European Research Council (ERC) via Consolidator Grant (724424 - No-LIMIT). PPB would like to thank Ministerio de Economía y Competitividad of Spain for his RyC contract.

# SUPPORTING INFORMATION
# From Dye Sensitized to Perovskite Solar Cells, The Missing Link

So-Min Yoo,[1,2] Seog Joon Yoon,[2,3] Juan A. Anta,[4] Hyo Joong Lee,[1,5]*
Pablo P. Boix,[6]* and Iván Mora-Seró[2]*

[1] Department of Chemistry, Chonbuk National University, Jeonju, 561–756 South Korea

[2] Institute of Advanced Materials (INAM), University Jaume I, Avenida de Vicent Sos Baynat, s/n, 12071 Castelló de la Plana, Spain

[3] Department of Chemistry, College of Natural Science, Yeungnam University, 280 Daehak-Ro, Gyeongsan, Gyeongbuk 38541, Republic of Korea

[4] Área de Química Física, Universidad Pablo de Olavide, Sevilla, Spain

[5] Department of Bioactive Material Sciences, Chonbuk National University, Jeonju, 561–756 South Korea

[6] Institut de Ciència Molecular, Universidad de València, C/J. Beltran 2, Paterna, Spain

* Corresponding author: solarlee@jbnu.ac.kr, pablo.p.boix@uv.es, sero@uji.es

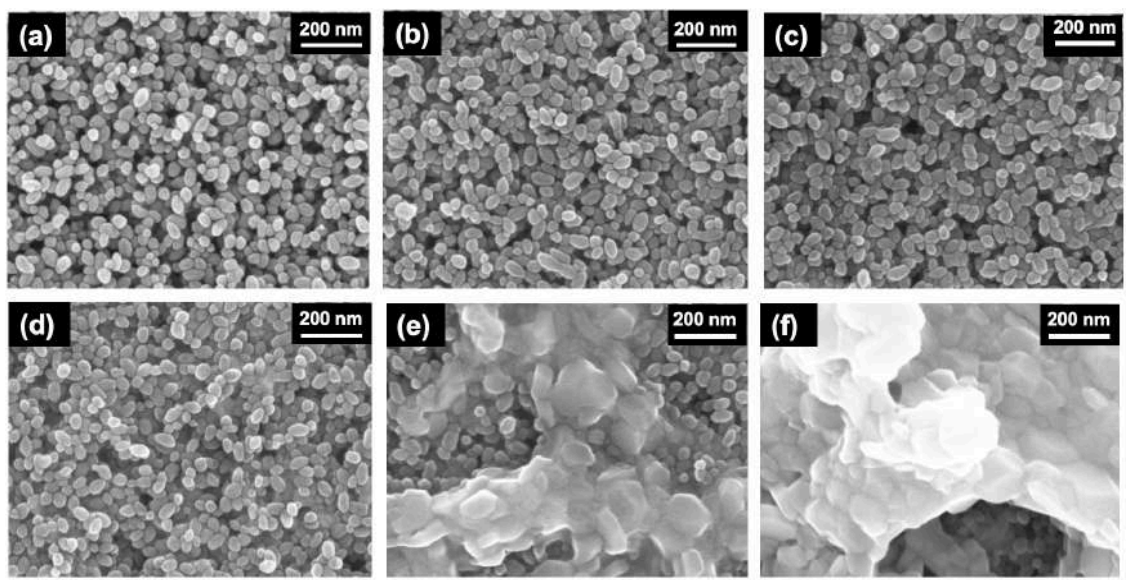

**Figure S1.** SEM micrographs of the top view of sensitized electrodes, corresponding to the cross sections plotted in Figure 1, using as precursor concentration of MAI 0.6 M and of $PbI_2$ of **(a)** 0.1 M; **(b)** 0.3 M; **(c)** 0.5 M; **(d)** 0.7 M; **(e)** 0.9 M and **(f)** 1.1 M.

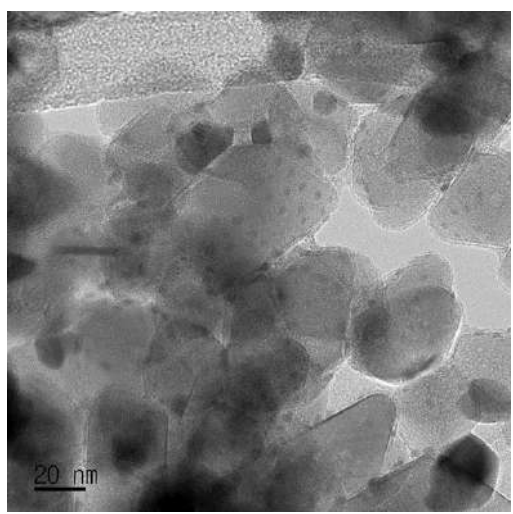

**Figure S2.** TEM images from 0.2 M PbI2 deposition and its subsequent formation of unconnected MAPbI$_3$ nanoparticles.

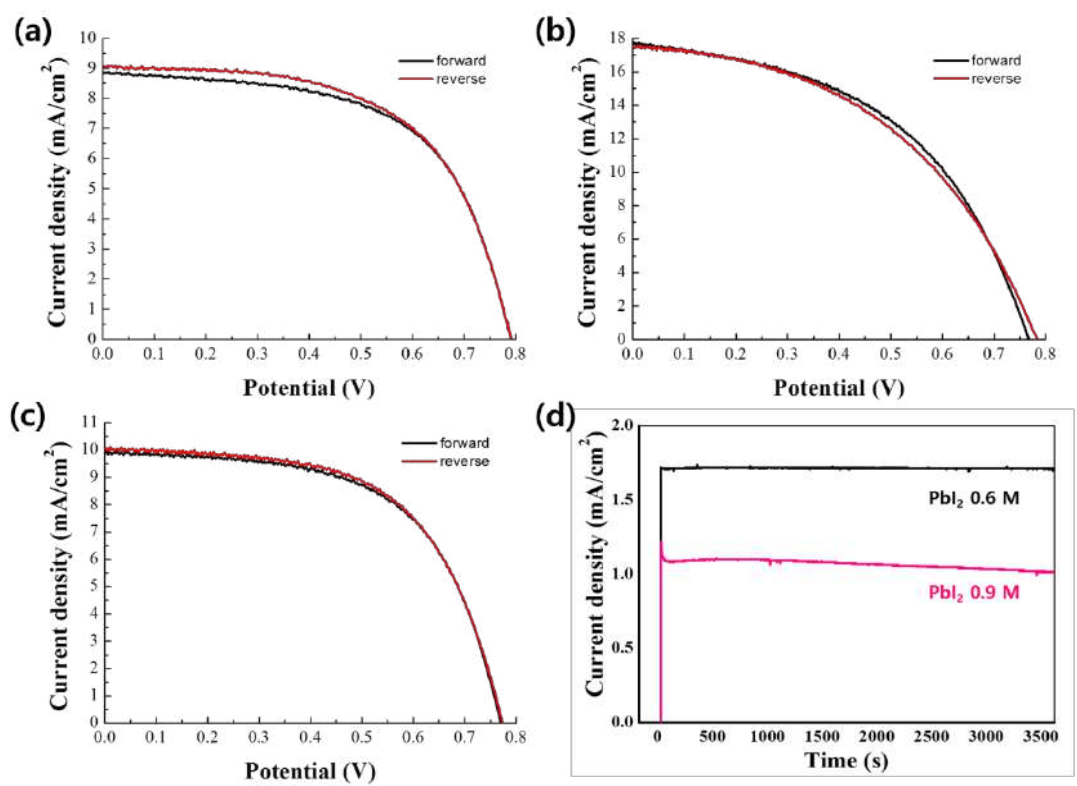

**Figure S3.** Example of hysteresis curves from cells prepared by using (a) 0.2 M, (b) 0.6 M, and 0.9 M PbI$_2$ at 1 sun illumination with a scan rate of 50 mV/sec. (d) Stabilized maximum power outputs were checked by measuring photocurrent at the applied V$_{max}$ under 0.1 sun illumination for 1 hour about two cells from 0.6 and 0.9 M PbI$_2$ deposition and its subsequent formation of MAPbI$_3$.

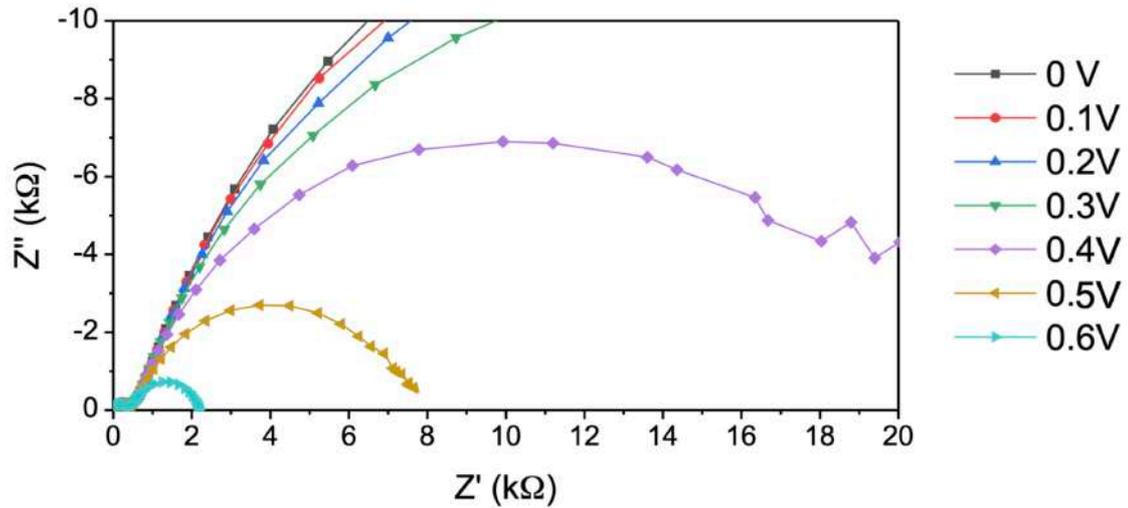

**Figure S4.** Nyquist plot of a MAPbI$_3$ solar cell using 0.2 M PbI$_2$ concentration at different applied bias under 0.1 sun illumination.

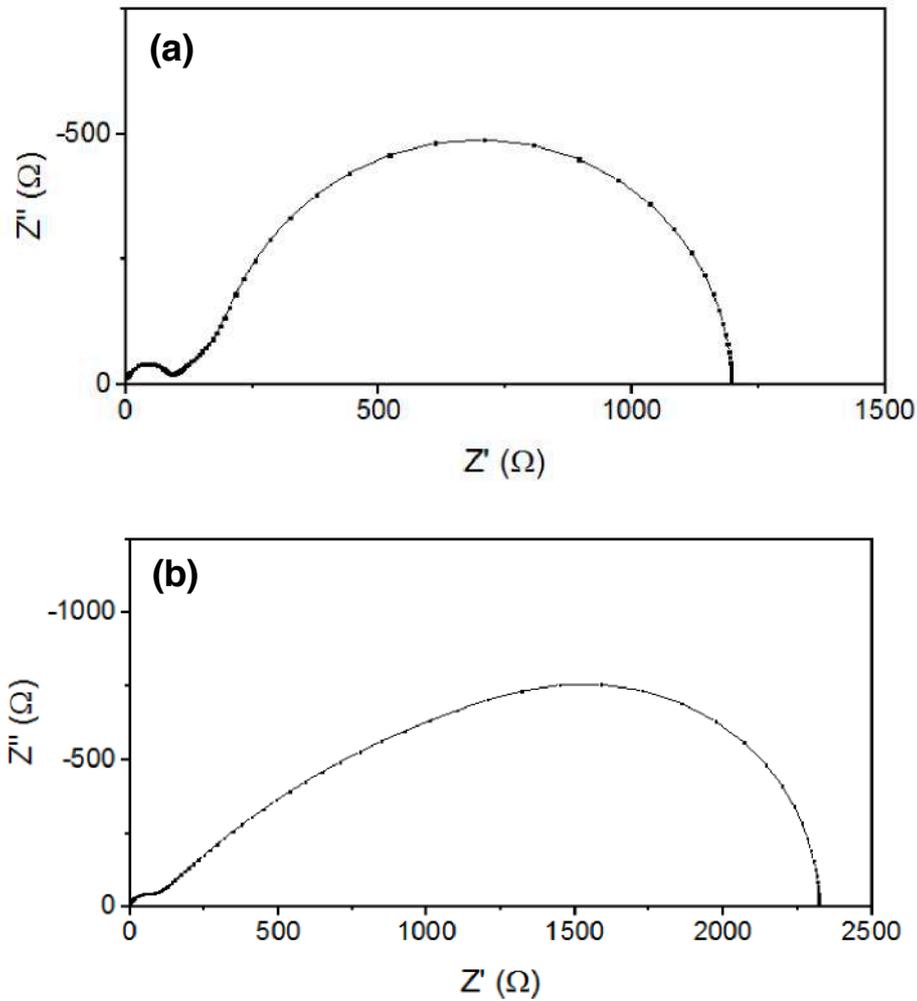

**Figure S5.** Simulations using the equivalent circuit plotted in Figure 3d of **(a)** the diffusion-recombination impedance, $R_{te}$=70 Ω, $R_{th}$=500 Ω, $R_{rec}$=1000 Ω, $C_\mu$=10$^{-5}$ F, $C_g$=10$^{-8}$ F, $R_{dr}$=7000 Ω, and $C_{dr}$=10$^{-6}$ F, and **(b)**

the Gerischer impedance, $R_{te}$=70 Ω, $R_{th}$=5000 Ω, $R_{rec}$=1000 Ω, $C\mu$=10$^{-5}$ F, $C_g$=10$^{-8}$ F, $R_{dr}$=7000 Ω, and $C_{dr}$=10$^{-6}$ F.

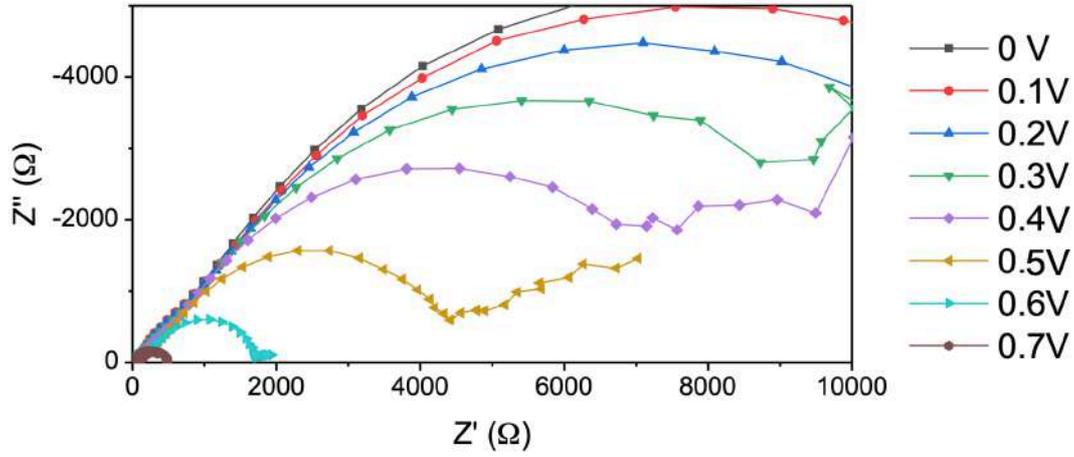

**Figure S6.** Nyquist plot of a MAPbI$_3$ solar cell using 0.6 M PbI$_2$ concentration at different applied bias under 0.1 sun illumination.

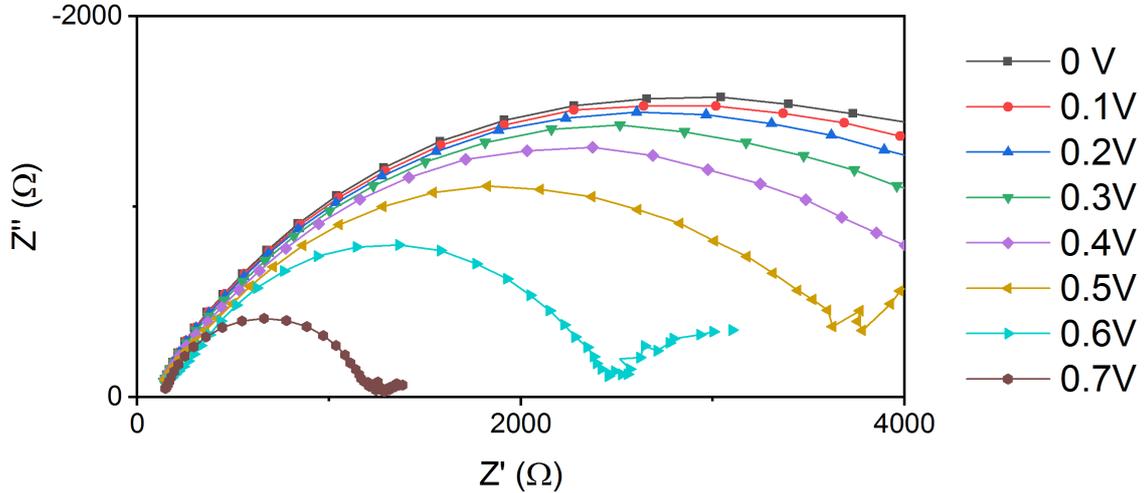

**Figure S7.** Nyquist plot of a MAPbI$_3$ solar cell using 0.9 M PbI$_2$ concentration at different applied bias under 0.1 sun illumination.

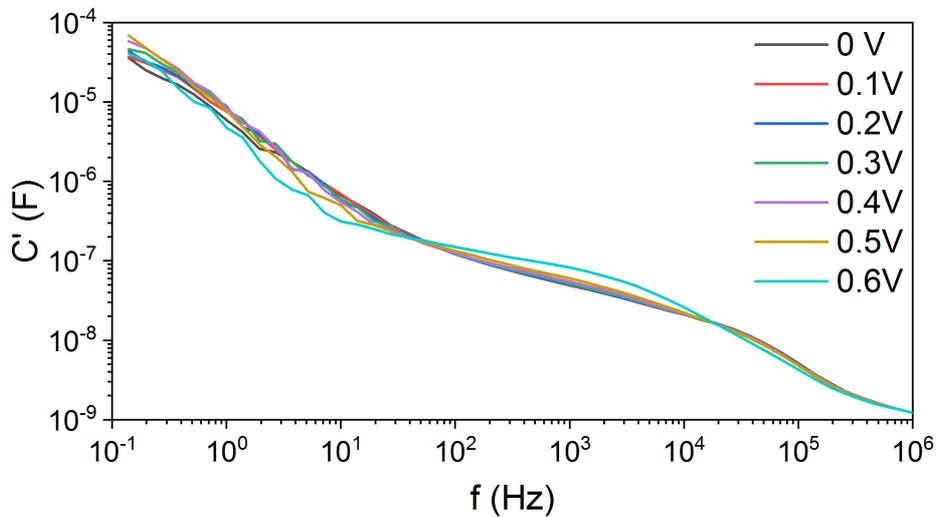

**Figure S8.** Bode plot of the real part of the capacitance of MAPbI$_3$ solar cells using 0.9 M PbI$_2$ concentration at different applied bias under 0.1 sun illumination.

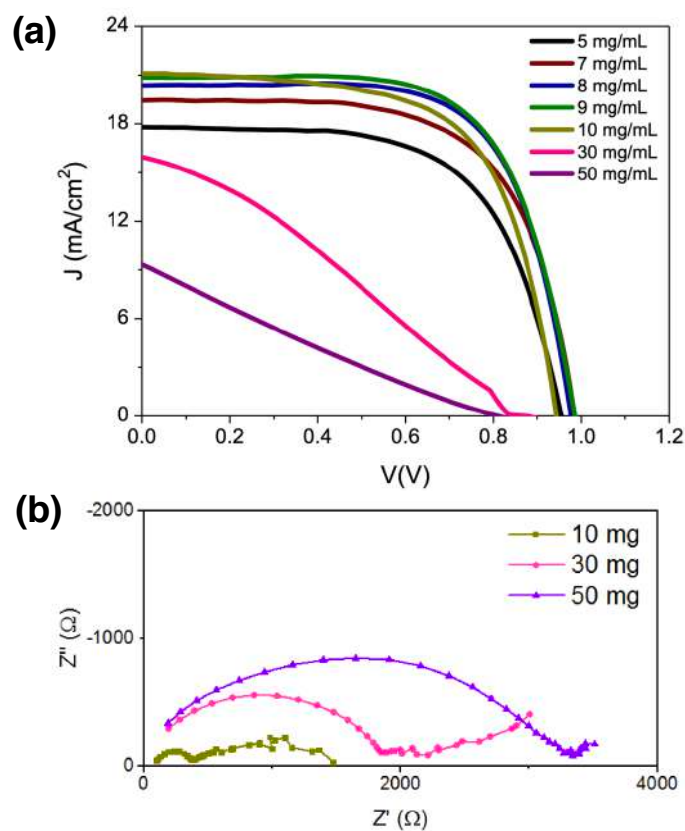

**Figure S9. (a)** Current potential curves of PSCs with n-i-p FU7 additive free as HTM.[1] For the deposition of FU7 different concentrations in chlorobenzene have been used, producing thicker HTM layers as the concentration increases. **(b)** Nyquist plot of PSCs using FU7 as HTM under 1 sun illumination at 0.4 V DC forward applied bias.

**Table S1.** Solar cell parameters of samples analyzed in Figure 2a, open circuit potential, $V_{oc}$, short circuit current, $J_{sc}$, fill factor, FF, and photoconversion efficiency, η.

|  | Jsc [mA cm$^{-2}$] | Voc [V] | FF [%] | η [%] |
|---|---|---|---|---|
| bare TiO$_2$ | - | - | - | - |
| 0.1 M PbI$_2$ | 4.1 | 0.721 | 57 | 1.6 |
| 0.3 M PbI$_2$ | 9.9 | 0.853 | 67 | 5.6 |
| 0.5 M PbI$_2$ | 12.4 | 0.830 | 60 | 6.3 |
| 0.7 M PbI$_2$ | 13.9 | 0.804 | 55 | 6.1 |
| 0.8 M PbI$_2$ | 11.8 | 0.820 | 58 | 5.6 |
| 0.9 M PbI$_2$ | 10.6 | 0.826 | 59 | 5.1 |
| 1.1 M PbI$_2$ | 10.9 | 0.831 | 58 | 5.3 |

**Note 1. Correlation with previous results.**

In this section, we try to discuss some of the previous results in the literature in terms of the model described in this work. One of the parameters more interesting that could be extracted from the impedance analysis is $R_{rec}$ as it allows to characterize an important part of the losses of the device. $R_{rec}$ can be defined generally as: [2, 3]

$$R_{rec} = R_{00} exp\left(-\frac{qV}{mk_BT}\right) \qquad (7)$$

where $R_{00}$ is the preexponential factor, $q$ the electron charge, $V$ the voltage, $m$ the ideality factor, $T$ the temperature and $k_B$ the Boltzmann constant. In a very recent work,[4] we suggest the identification of a resistance linearly proportional to $R_{rec}$ by the determination of the ideality factor from the slope of $\log(R)$ vs. V using the Eq. (7), where $R$ is obtained from IS, and comparing the extracted $m$ with the ideality factor obtained by the analysis of $V_{OC}$ at different light illumination, $\phi$, by the relation:[4, 5]

$$V_{oc} = \frac{E_g}{q} - \frac{mk_BT}{q}ln(\phi) \qquad (8)$$

where $E_g$ is the energy of the bandgap.

On the other hand, the slope correlation of $\log(R_{HF})$ and $\log(R_{LF})$ vs. $V_{oc}$ has been also studied in previous works. Some works reported similar slopes for $R_{HF}$ and $R_{LF}$, fact that was explained suggesting that both arcs are referred to the same physical phenomena.[6-8] In contrast, other works observed different slopes for $R_{HF}$ and $R_{LF}$.[3, 4, 9] Our model could help to understand both situations. In detail, Eqs (3) and (4) can be rewritten using Eq. (5) (Eqs. 3-5 in the main text) as:

$$R_{HF} = (R_{rec} + R_t/3)\frac{1}{1+\frac{R_{LF}}{R_{HF}}} \qquad (9)$$

$$R_{LF} = (R_{rec} + R_t/3)\frac{\frac{R_{LF}}{R_{HF}}}{1+\frac{R_{LF}}{R_{HF}}} \qquad (10)$$

When $R_{LF}/R_{HF}$ ratio presents a weak voltage dependence in comparison with the exponential dependence of $R_{rec}$, see Eq. (7), similar slope dependence is obtained for $R_{LF}$ and $R_{HF}$ as observed in previous works.[6, 7] Consequently, the common physical phenomena for both arcs suggested by the authors of previous work should be recombination (coupled with transport if it is not negligible). Note that this situation could be accomplished if $R_{dr}$ presents an exponential dependence similar to $R_{rec}$ while $R_t$ exhibits weaker voltage dependence (ohmic resistance fulfill this condition).[6, 10] However, a stronger $R_{LF}/R_{HF}$ ratio dependence with voltage can produce different $R_{HF}$ and $R_{LF}$ slopes, as it has been also reported.[3, 4] To this extend, the identification of the exact physical processes accounting for the dielectric relaxation-like branch would represent another step towards a full interpretation of the IS spectra of PSCs, providing additional information of the working mechanism. The extended IS data compilation, more detailed temperature measurements and the utilization of complementary characterization techniques will undoubtedly pave the way for the characterization of these dielectric relaxation-like processes. As we have already commented this study is beyond the scope of the present manuscript and it is receiving an increasing attention in the literature.[8, 10-13]